\documentclass[aps,prl,reprint,superscriptaddress]{revtex4-1}

\usepackage{amsmath,graphicx,color}

		\newcommand{\rr}[0]{\boldsymbol{r}}
		\newcommand{\qq}[0]{\boldsymbol{q}}
		\newcommand{\VV}[0]{\boldsymbol{V}}
		
		\newcommand{\Dc}[0]{D_\mathrm{p}}

\begin{document}


\title{Active phase separation in mixtures of chemically interacting particles}


\author{Jaime Agudo-Canalejo}
\email{jaime.agudocanalejo@physics.ox.ac.uk}
\affiliation{Rudolf Peierls Centre for Theoretical Physics, University of Oxford, Oxford OX1 3PU, United Kingdom}
\affiliation{Department of Chemistry, The Pennsylvania State University, University Park, Pennsylvania 16802, USA}

\author{Ramin Golestanian}
\email{ramin.golestanian@ds.mpg.de}
\affiliation{Max Planck Institute for Dynamics and Self-Organization (MPIDS), D-37077 G\"ottingen, Germany}
\affiliation{Rudolf Peierls Centre for Theoretical Physics, University of Oxford, Oxford OX1 3PU, United Kingdom}


\date{\today}

\begin{abstract}
We theoretically study mixtures of chemically-interacting particles, which produce or consume a chemical to which they are attracted or repelled, in the most general case of many coexisting species. We find a new class of active phase separation phenomena in which the nonequilibrium chemical interactions between particles, which break action-reaction symmetry, can lead to separation into phases with distinct density and stoichiometry. Due to the generic nature of our minimal model, our results shed light on the underlying fundamental principles behind nonequilibrium self-organization of cells and bacteria, catalytic enzymes, or phoretic colloids.
\end{abstract}


\maketitle

Microorganisms and cells can chemotax in response to gradients of chemicals that they themselves produce or consume \cite{wadh04,vanh04}. The same behaviour has been recently observed at the nanoscale for individual enzymes \cite{seng13,jee17,agud18a}, and can be mimicked in synthetic systems using catalytically-active phoretic colloids  \cite{ande89,soto14,soto15,niu17a,niu18,yu18,varm18,rall19} . Importantly, when many such particles are placed in solution, they interact with each other through their influence on the chemical's concentration field. Chemical interactions underlie a wide variety of phenomena such as self-organisation in heterogeneous populations of microorganisms and cells (e.g. quorum sensing \cite{kell06} and competition for nutrients \cite{hibb10} in bacterial ecosystems, or cell-cell communication \emph{via} chemokines \cite{frie09}); aggregation of enzymes that participate in common catalytic pathways into a metabolon \cite{swee18,wu15,zhao17}, which could be harnessed in the design of better synthetic pathways \cite{schw16}; or the self-assembly of active materials from catalytic colloids \cite{pala13,mass18}.

A key feature of chemical interactions between two different species---whether they are synthetic catalytic colloids, biological enzymes, or whole cells or microorganisms---is that they are in general non-reciprocal \cite{soto14,soto15}. The concentration field of a fast-diffusing chemical around a chemically-active particle of species $i$ is, to lowest order, given by $c-c_0 \propto \alpha_i / r$ where $\alpha_i$ is the activity of the species (positive and negative for producer and consumer species), $r$ is the distance to the particle's centre, and $c_0$ is the reference concentration of the chemical at infinity. In turn, the motion of a particle of species $j$ in response to gradients of the chemical is given by a velocity $\VV_j = - \mu_j \nabla c$ where $\mu_j$ is the mobility of the species (positive or negative if the species is directed towards regions of lower or higher concentration of the chemical). Combining these two expressions, one finds that the velocity of the $j$-species particle in response to the presence of the $i$-species particle is $\VV_{ij} \propto \alpha_i \mu_j \rr_{ij} / |\rr_{ij}|^3$ with $\rr_{ij} = \rr_{j} - \rr_{i}$, whereas the velocity of the latter in response to the presence of the former is $\VV_{ji} \propto - \alpha_j \mu_i \rr_{ij} / |\rr_{ij}|^3$. Note that in general $\VV_{ij} \neq - \VV_{ji}$ for $i \neq j$ because $\alpha_i \mu_j \neq \alpha_j \mu_i$, implying a broken action-reaction symmetry for inter-species interactions, which would be impossible in a system at thermodynamic equilibrium \cite{ivl15}. 

We have performed Brownian dynamics simulations \cite{Note1} of the model just described for a wide range of mixtures of chemically-interacting species; see Fig.~\ref{intro} and movies 1--12 in the Supplemental Material \cite{Note1}. For binary mixtures we find that, while in a large region of the parameter space the mixtures remain homogeneous, the homogeneous state can also become unstable leading to a great variety of phase separation phenomena. Here, phase separation is used in the sense of macroscopic (system-spanning) separation typically into a single large cluster (occasionally into two; see Fig.~\ref{intro}(a)) that coexists with a dilute (or empty) phase. The phase separation process may lead to aggregation of the two species into a single mixed cluster, or to separation of the two into either two distinct clusters or into a cluster of a given stoichiometry and a dilute phase. The resulting configurations are qualitatively distinct for mixtures of one chemical-producer and one chemical-consumer species, as opposed to mixtures of two producer (or consumer) species; compare panels (a) and (c) in Fig.~\ref{intro}. While the typical steady-state configurations are static, for mixtures of producer and consumer species we also find that static clusters can undergo a shape-instability that breaks their symmetry, leading to a self-propelling macrocluster (Fig.~\ref{intro}(b), movies 8 and 9). Randomly-generated highly-polydisperse mixtures of up to 20 species also show homogeneous as well as phase-separated states (Fig.~\ref{intro}(d), movies 11 and 12).

 \begin{figure*}
 \includegraphics[width=1\linewidth]{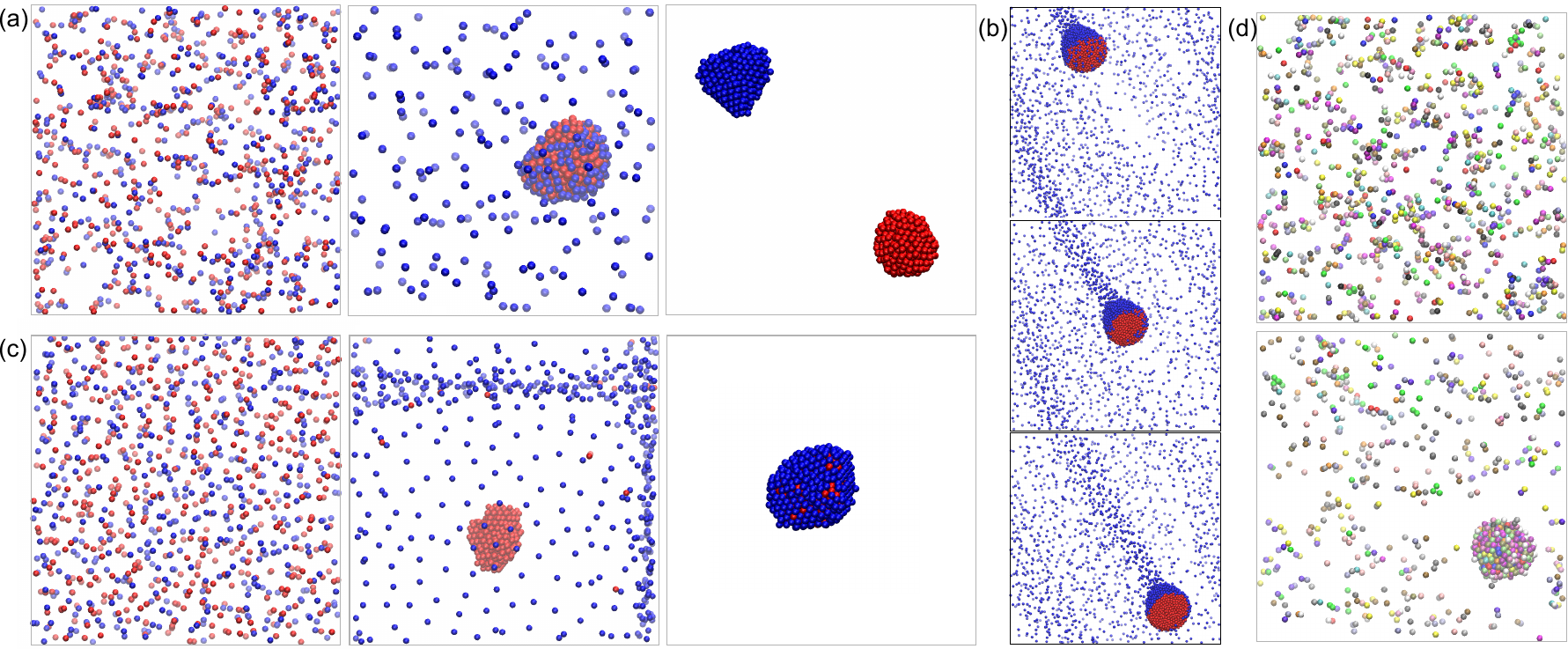}
 \caption{Mixtures of chemically-interacting particles display a wealth of active phase separation phenomena. (a) Binary mixtures of producer (blue) and consumer (red) species show, from left to right, homogeneous states with association of particles into small ``molecules'', aggregation into a static dense phase that coexists with a dilute phase, and separation into two static collapsed clusters; see movies 1--3  in the Supplemental Material \cite{Note1}. (b) The static aggregate (a, centre) can undergo symmetry breaking to form a self propelled macroscopic cluster; see  movies 8 and 9. (c) Binary mixtures of producer species (blue and red) show homogeneous states without molecule formation, separation into a static dense phase and a dilute phase that  is depleted near the dense phase, and aggregation into a static collapsed cluster; see  movies 4--6. (d) Randomly-generated highly polydisperse mixtures (20 different species) can remain homogeneous or undergo macroscopic phase separation; see movies 11 and 12. Simulation parameters for each case (a--c) can be found in the description of the corresponding  movies. \label{intro}}
 \end{figure*}

In the following, we will show how these results can be understood by means of a continuum theory, and how the observed phase separation behaviour is intimately related to the non-equilibrium and non-reciprocal character of the interactions. This represents a fundamentally new class of active phase separation, in which the activity arises from the non-equilibrium nature of the interactions between particles that are otherwise non-motile, rather than from the intrinsic activity of self-propelling particles as commonly studied \cite{vics95,tone98,gole12,fily12,redn13,saha14,cohe14,zott14,pohl14,cate15,lieb15,lieb17}.

We consider a system consisting of $M$ different species of chemically-interacting particles, with concentrations $\rho_i(\rr,t)$ for $i=1,...,M$; and a messenger chemical with concentration $c(\rr,t)$. The concentration of species $i$ is described by $\partial_t \rho_i(\rr,t) - \nabla \cdot [\Dc \nabla \rho_i + (\mu_i \nabla c) \rho_i ]=0$ which includes a diffusive term with diffusion coefficient $\Dc$, which for simplicity is taken to be equal for all species (implying that all particles are of similar size or, in the case of microorganisms, all species show a similar baseline level of non-directed random motion); as well as an advective term describing motion in response to gradients of the chemical. The concentration of the chemical is described by $\partial_t c(\rr,t) - D \nabla^2 c = \sum_{i} \alpha_i \rho_i$ which includes diffusion with coefficient $D$, and production or consumption of the chemical by all particle species. Performing a linear stability analysis \cite{Note1} of this coupled system of $M+1$ equations around a spatially-homogeneous state with particle densities $\rho_i(\rr,t)=\rho_{0i}$, in the limit of a fast-diffusing chemical, we find that the homogeneous state becomes unstable towards a spatially-inhomogeneous state when the following condition holds
\begin{equation}
\sum_i \mu_i \alpha_i \rho_{0i} < 0.
\label{instNsame}
\end{equation}
The instability corresponds to macroscopic phase separation, in the sense that it occurs for perturbations of infinite wave length, specifically for perturbations with wave number $\qq^2 < -  (D \Dc)^{-1} \sum_i \mu_i \alpha_i \rho_{0i}$, with those having infinite wave length $\qq \to 0$ being the first and most unstable. Importantly, the stability analysis also tells us about the stoichiometry of the different particle species at the onset of growth of the perturbation, which follows
\begin{equation}
(\delta \rho_1,\delta \rho_2,...,\delta \rho_M) = \left(1,\frac{\mu_2 \rho_{02}}{\mu_1 \rho_{01}},...,\frac{\mu_M \rho_{0M}}{\mu_1 \rho_{01}}\right)\delta \rho_1.
\label{stoichNsame}
\end{equation}

 \begin{figure*}
 \includegraphics[width=1\linewidth]{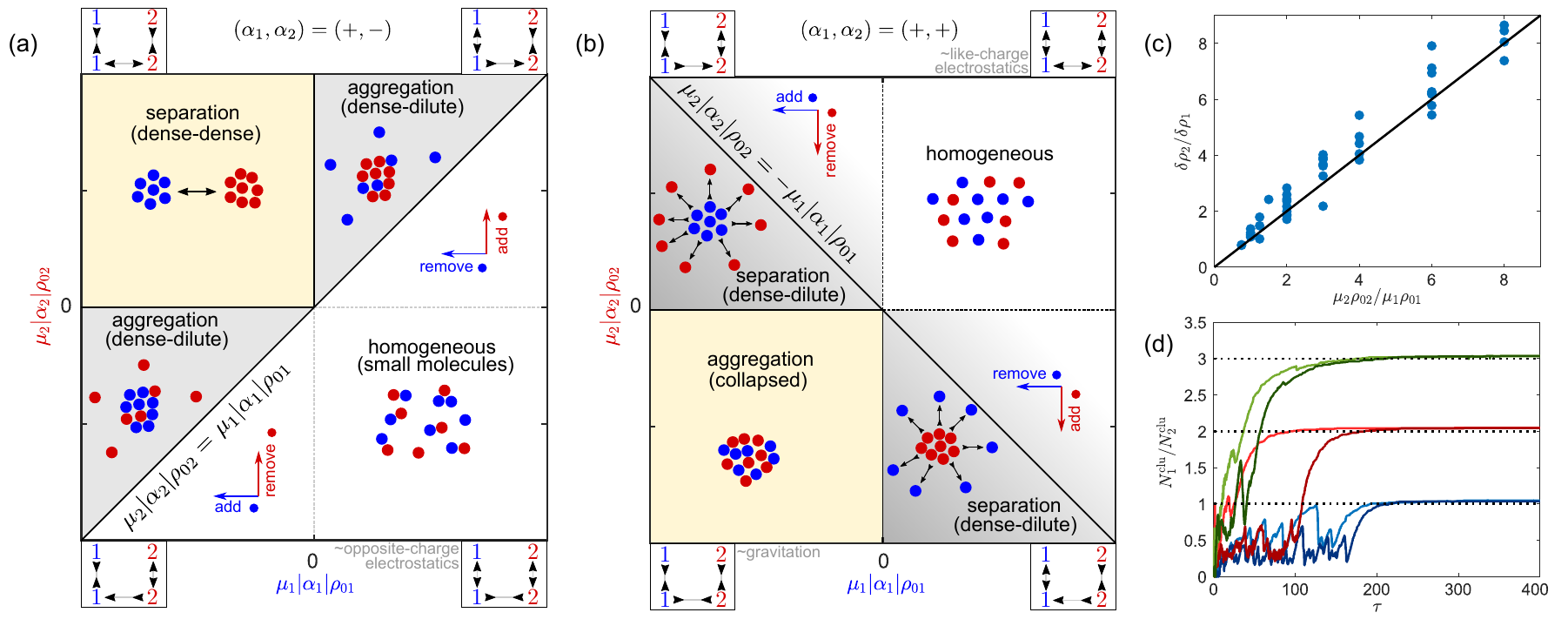}
 \caption{(a) Stability diagram for mixtures of one producer and one consumer species (cf.~Fig.~\ref{intro}(a)), and (b) for mixtures of two producer species (cf.~Fig.~\ref{intro}(c)). In (a,b) the boxed legends attached to each quadrant symbolise the ``interaction network'' representing the sign of interactions between each species in the system, as described in the main text. Phase separation (aggregation in (a), separation in (b)) can be triggered by addition or removal of particles (density changes) only when interactions between the two species are intrinsically non-reciprocal. (c) Stoichiometry at the onset of the instability, obtained from 44 simulations (blue circles, see Table~S1 in the Supplemental Material \cite{Note1}) compared to the stability analysis prediction (Eq.~\ref{stoichNsame}). (d) Time evolution of the stoichiometry of the biggest cluster arising from aggregation of $(\alpha_1,\alpha_2)=(+,-)$ mixtures, demonstrating that the long time stoichiometry is predicted by the neutrality rule (Eq.~\ref{neutr}) and is independent of the species' mobility (blue: $\tilde{\alpha}_2=-1$, $\tilde{\mu}_2=8$ and $12$; red: $\tilde{\alpha}_2=-2$, $\tilde{\mu}_2=4$ and $8$; green: $\tilde{\alpha}_2=-3$, $\tilde{\mu}_2=3$ and $5$; in all cases $N_1=800$, $N_2=200$, $\tilde{\alpha}_1=\tilde{\mu}_1=1$). \label{phasediag}}
 \end{figure*}

If only a single particle species is present ($M=1$), the instability criterion (\ref{instNsame}) describes the well-known Keller-Segel instability \cite{kell70}, which simply says that the homogeneous state is stable for particles that repel each other ($\mu_1 \alpha_1>0$), whereas particles that attract each other ($\mu_1 \alpha_1<0$) tend to aggregate, with the end state being a featureless macroscopic cluster containing all particles. In contrast, we will now show that as soon as we have mixtures of more than one species, the combination of the instability criterion (\ref{instNsame}) and the stoichiometric relation (\ref{stoichNsame}) predicts a wealth of new phase separation phenomena.

\begin{figure}[t]
 \includegraphics[width=1\linewidth]{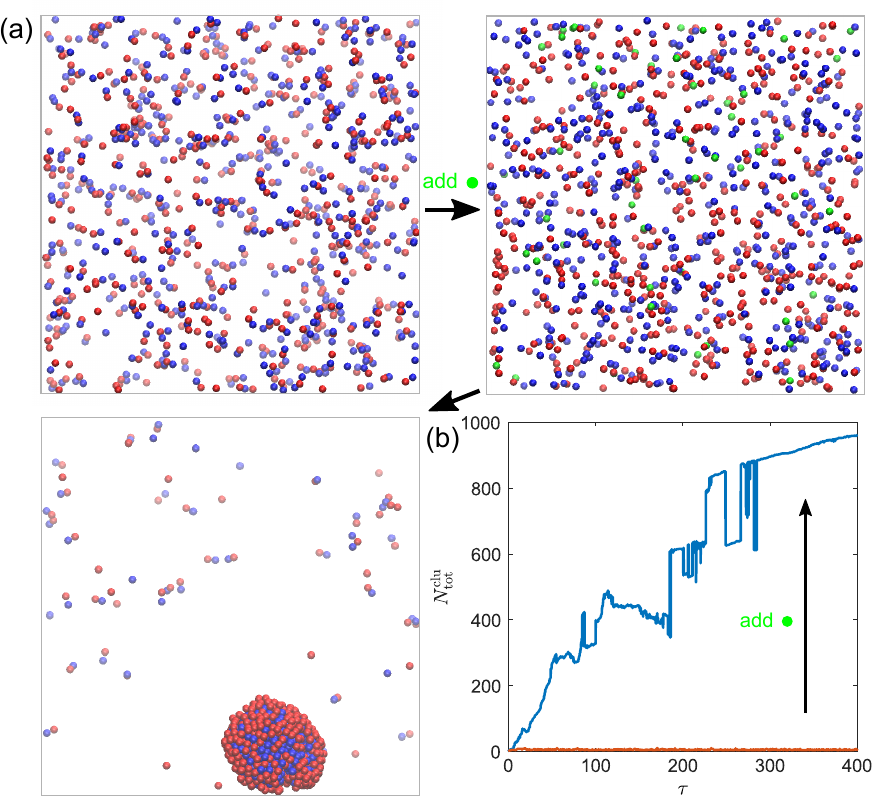}
 \caption{Phase separation induced by a small amount of an active ``doping agent''. (a) Simulation snapshots showing macroscopic aggregation of a previously homogeneous mixture ($N_1=N_2=500$, $\tilde{\alpha}_1=\tilde{\mu}_1=1$, $\tilde{\alpha}_2 = -1$, $\tilde{\mu}_2=1/2$) after addition of 5 \% of a third species ($N_3=50$, $\tilde{\alpha}_3=-5$, $\tilde{\mu}_3=2$), compare movies 1 and 10  in the Supplemental Material \cite{Note1}. (b) Time evolution of the size of the largest cluster (total number of particles), in the absence and presence of the third species. \label{dope}}
 \end{figure}

For binary mixtures ($M=2$), the instability condition (\ref{instNsame}) becomes $\mu_1 \alpha_1 \rho_{01} + \mu_2 \alpha_2 \rho_{02} < 0$, and the stoichiometric constraint (\ref{stoichNsame}) implies that when $\mu_1$ and $\mu_2$ have equal or opposite sign, the instability will lead respectively to aggregation or separation of the two species. Combining these criteria we can construct a stability diagram for the binary mixture, although we must distinguish between two qualitatively-different kinds of mixtures: those of one producer and one consumer species, see Fig~\ref{phasediag}(a) where we have chosen $(\alpha_1,\alpha_2)=(+,-)$ without loss of generality; and those of two producer species, see Fig~\ref{phasediag}(b). The case of two consumer species is related to the latter by the symmetry $(\mu_1,\mu_2) \to -(\mu_1,\mu_2)$; see Fig.~S1  in the Supplemental Material \cite{Note1}. In this way, the parameter space for each type of mixture can be divided into regions leading to homogeneous, aggregated, or separated states, which correspond directly to those observed in simulations; compare Fig~\ref{phasediag}(a,b) to Fig.~\ref{intro}(a,c). We note, however, that while for $(\alpha_1,\alpha_2)=(+,-)$ mixtures the simulations are always seen to match the predicted phase behaviour, for $(\alpha_1,\alpha_2)=(+,+)$ mixtures we have observed separation even when the continuum theory predicts the homogeneous state to be linearly stable, although proceeding much more slowly (see movie 7  in the Supplemental Material \cite{Note1}), indicating that in this region separation may be occurring through a nucleation-and-growth process controlled by fluctuations. This is denoted as the shaded gray region extending past the instability line in Fig~\ref{phasediag}(b).

The wide variety of phase separation phenomena arising in these mixtures is intimately related to the active, non-reciprocal character of the chemical interactions. In particular, it is useful to consider the sign of both inter-species as well as intra-species interactions (as described above, species $i$ is attracted to or repelled from species $j$ when $\mu_i \alpha_j$ is negative or positive, respectively). In the stability diagrams in Fig.~\ref{phasediag}(a,b), we find that each quadrant corresponds to a distinct ``interaction network'' between species, as depicted in the boxed legends attached to every quadrant (as an example, the top-right interaction network in (a) can be read as ``1 is attracted to 2, 2 is repelled from 1, 1 is repelled from 1, and 2 is attracted to 2''). We find that only three regions in the parameter space have passive analogs: (i) The bottom-right of (a) corresponds to electrostatics with opposite charges, where equals repel and opposites attract, allowing for the formation of small active molecules as studied in Refs.~\citenum{soto14} and \citenum{soto15}. (ii) The top-right of (b) corresponds to electrostatics with like charges, where all interactions are repulsive leading to a homogeneous state. (iii) The bottom-left of (b) corresponds to gravitation, where all interactions are attractive. The top-left of (a) can be thought of as the opposite of electrostatics (or as gravitation including a negative mass species), where equals attract and opposites repel. The remaining four quadrants involve intrinsically non-reciprocal interactions where one species chases after the other: in (a), a self-repelling species chases after a self-attracting species; whereas in (b), a self-attracting species chases after a self-repelling species. Importantly, we observe that the most non-trivial instances of phase separation, which are also those that can be triggered simply by density changes (e.g. by addition or removal of particles), occur in regions with such chasing interactions, which are in turn a direct signature of non-equilibrium activity. 

Fourier analysis \cite{Note1} of the simulation results (44 simulations with varying $N_i$, $\alpha_i$, and $\mu_i$; see Table~S1  in the Supplemental Material \cite{Note1}) agrees quantitatively with the theoretical prediction (\ref{stoichNsame}) for the stoichiometry at the onset of the instability; see Fig.~\ref{phasediag}(c). However, this initial value is not representative of the long-time stoichiometry of the phases. For $(\alpha_1,\alpha_2)=(+,+)$ mixtures, shown in Figs.~\ref{intro}(c) and \ref{phasediag}(b), we always observe final configurations with either complete aggregation or separation of the two species. For $(\alpha_1,\alpha_2)=(+,-)$ mixtures, shown in Figs.~\ref{intro}(a) and \ref{phasediag}(a), we always observe complete separation, but aggregation in this case leads to a cluster with non-trivial stoichiometry (Fig.~\ref{intro}(a), centre). Phenomenologically, we observe that the formation of such clusters proceeds by fast initial aggregation of the particles of the self-attractive species ($\alpha_i \mu_i<0$) followed by slower recruitment of particles of the self-repelling species ($\alpha_i \mu_i>0$) until the cluster is chemically ``neutral'', in the sense that its net consumption or production of chemicals vanishes, namely
\begin{equation}
\alpha_1 N_1^\mathrm{clu}+\alpha_2 N_2^\mathrm{clu} = 0,
\label{neutr}
\end{equation}
where $N_i^\mathrm{clu}$ is the number of particles of species $i$ in the cluster. The long-time stoichiometry of the clusters thus depends on the activity of the species, but it is independent of their mobility; see Fig.~\ref{phasediag}(d). An intuitive explanation for this observation can be provided as follows: once the cluster becomes neutral, the remaining self-repelling particles will no longer ``sense'' its presence and stay in a dilute phase. However, at high values of activity and mobility for the self-attractive species, deep inside the instability region, these static neutral clusters can become unstable \emph{via} shape-symmetry breaking towards a self-propelled asymmetric cluster (Fig.~\ref{intro}(b)), which also involves the ``shedding'' of some of the self-repelling particles; see Fig.~S3 and movies 8 and 9  in the Supplemental Material \cite{Note1}. Finding a precise criterion for this symmetry-breaking to occur remains an open question, but we note that the existence of self-propelled clusters is a clear sign of non-equilibrium physics.

Going beyond binary mixtures ($M>2$), the phase separation phenomenology becomes even more complex due to the increasing number of parameter combinations, leading to a large variety of possible interaction networks between the different species. The instability condition (\ref{instNsame}) remains extremely useful, however. As a first example, in Fig.~\ref{dope} we demonstrate how a small amount of a highly active ``dopant'' third species can be added to an otherwise homogeneous binary mixture in order to trigger macroscopic phase separation of the whole mixture on demand; see also  movie 10 in the Supplemental Material \cite{Note1}. As a second example, we have simulated highly polydisperse mixtures made up of 20 different species with activities and mobilities randomly chosen in the intervals $-2 \leq \tilde{\alpha}_i,\tilde{\mu}_i \leq +2$ for each species; see Fig.~\ref{intro}(d) and movies 11 and 12. We find that the instability criterion (\ref{instNsame}) can rather reliably distinguish between phase-separating and homogeneous mixtures; see Fig.~S4 in the Supplemental Material \cite{Note1}. We note that while all mixtures we predicted to phase-separate did so, some mixtures for which we predicted a linearly-stable homogeneous state were observed to phase-separate, albeit more slowly, once again pointing to a nucleation-and-growth mechanism rather than to a linear instability.

We have presented here a minimal model for phase separation in mixtures of chemically-interacting particles, and the generic phenomena that we predict should be applicable to a wide variety of systems. In the context of morphogenesis and collective migration in bacterial colonies and cells in tissues, the prediction of a transition between static and self-propelled clusters is particularly interesting. Here, it is important to take into account that what we call here ``two species'' may also represent a single species in two distinct states, each with different chemical activity or chemotactic behaviour. Regarding metabolon formation by enzymes in catalytic pathways, our prediction of ``neutral'' clusters (Eq.~\ref{neutr}) is most intriguing, as it would correspond to a cluster in which one enzyme channels all of its product to be taken as substrate by the next enzyme, with no substrate missing or in excess. Finally, our predictions can be tested in detail in experiments using synthetic catalytic colloids, by systematically varying the sign and magnitude of the chemical activity, as well as the concentration of the different species. In future work, it will be interesting to characterize in more detail the non-equilibrium activity of the system by means of its energy dissipation or entropy production \cite{saba12,gasp18}. Moreover, we note that in our simulations we have neglected hydrodynamic interactions between particles as well as near-field contributions in the chemical concentrations \cite{Note1}. While we we do not expect our results for the onset and stoichiometry of the instability to change, the detailed dynamics of aggregation and growth of the clusters as well as their internal dynamics will be affected by these additional effects.

This study was supported by the US National Science Foundation under MRSEC Grant number DMR-1420620.

\nocite{stra99,touk03}

%

\end{document}



\title{Supplemental Material -- Active phase separation in mixtures of chemically interacting particles}


\author{Jaime Agudo-Canalejo}
\email{jaime.agudocanalejo@physics.ox.ac.uk}
\affiliation{Rudolf Peierls Centre for Theoretical Physics, University of Oxford, Oxford OX1 3PU, United Kingdom}
\affiliation{Department of Chemistry, The Pennsylvania State University, University Park, Pennsylvania 16802, USA}

\author{Ramin Golestanian}
\email{ramin.golestanian@ds.mpg.de}
\affiliation{Max Planck Institute for Dynamics and Self-Organization (MPIDS), D-37077 G\"ottingen, Germany}
\affiliation{Rudolf Peierls Centre for Theoretical Physics, University of Oxford, Oxford OX1 3PU, United Kingdom}


\date{\today}



\maketitle

\renewcommand\thefigure{S\arabic{figure}}    
\renewcommand\thetable{S\arabic{table}}  
\renewcommand\theequation{S\arabic{equation}}

\section{Linear stability analysis}

We consider small deviations from a homogenous state, so that the particle density is described by $\rho_i(\rr,t) = \rho_{0i} + \delta \rho_i(\rr,t)$. The net catalytic activity of a mixture is defined as $A \equiv \sum_{i} \alpha_i \rho_{0i}$, where we note that $A$ represents activity in the homogeneous state, while locally we have $\sum_{i} \alpha_i \rho_{i} = A + \sum_{i} \alpha_i \delta \rho_{i}$. The chemical concentration can be separated into a (time-dependent) uniform value and the deviations from this uniform value in response to nonuniformities of the particle distribution, so that $c(\rr,t) = c_0 + A t + \delta c (\rr, t)$. Introducing this into the evolution equation for $c(\rr,t)$ we obtain an equation for the deviations $\delta c (\rr,t)$ given by
\begin{equation}
\partial_t \delta c(\rr,t) - D \nabla^2 \delta c = \sum_{i} \alpha_i \delta \rho_i.
\label{deltac}
\end{equation}
Because the small chemical diffuses much faster than the large particles, the deviations $\delta c(\rr,t)$ of the chemical concentration from the uniform value $c_0+At$ can be assumed to reach a steady state instantaneously for each configuration of the particles, so that from (\ref{deltac}) we obtain $- D \nabla^2 \delta c =  \sum_{i} \alpha_i \delta \rho_i$. Introducing this into the evolution equation for $\rho_i(\rr,t)$, and staying only to linear order in $\delta \rho_i(\rr,t)$, we obtain
\begin{equation}
\partial_t \delta \rho_i(\rr,t) = \Dc \nabla^2 \delta \rho_i - \frac{\mu_i \rho_{0i}}{D} \sum_{j} \alpha_j \delta \rho_j.
\label{deltarho}
\end{equation}

The linearised system of equations (\ref{deltarho}) with $i=1,...,M$ describes the evolution of the deviations of the particle density around the homogeneous state.  We have assumed above that there is no external input or removal of chemicals into or out of the solution. In this case, our results will be valid for mixtures with net positive production ($A > 0$) as long as there is still a high enough concentration of whatever precursor substance is needed for the production of the chemical; and for mixtures with net consumption ($A<0$) as long as the concentration of the chemical is still high enough. More precisely, the relevant concentrations should be high enough that the production or consumption of chemical by the particles can be considered to be taking place in the saturated regime, i.e. at a rate $\alpha_i$ that is concentration-independent. Balanced mixtures with no net production or consumption ($A=0$) could in principle be kept active indefinitely as long as there is an external energy source (e.g. light), necessary for the cyclic transformation of the chemical. Lastly, in all cases one could externally supply the necessary chemicals to keep the reactions active. In the absence of external input, all mixtures will eventually reach chemical equilibrium and activity will stop.

The stability analysis of (\ref{deltarho}) is done most conveniently by defining the new variables $u_i \equiv \alpha_i \delta \rho_i$ and the parameters $\gamma_i \equiv \mu_i \alpha_i \rho_{0i}/D$. The system of equations (\ref{deltarho}) can be rewritten as
\begin{equation}
[\partial_t - \Dc \nabla^2 + \gamma_i ] u_i + \gamma_i \sum_{j \neq i} u_j = 0
\label{u1}
\end{equation}
the solution of which is given by a sum of Fourier modes of the form $u_i (\rr,t) = u_{\qq i} \ex{\ii \qq \cdot \rr} \ex{\lambda t}$. Introducing this into (\ref{u1}) finally results in the eigenvalue problem
\begin{equation}
[\lambda + \Dc \qq^2 + \gamma_i ] u_{\qq i} + \gamma_i \sum_{j \neq i} u_{\qq j} = 0
\label{u2}
\end{equation}
for the growth rate $\lambda$ of the perturbation modes with wavenumber $\qq$.

By defining $\tilde{\lambda} \equiv - (\lambda + \Dc \qq^2)$, the eigenvalue problem (\ref{u2}) is equivalent to finding the eigenvalues of a $M \times M$ matrix with $M$ identical rows each given by $[\gamma_1~\gamma_2~...~\gamma_M]$. Such a matrix has rank 1 and therefore at least $M-1$ of its eigenvalues are equal to zero, $\tilde{\lambda}_-=0$. Because the trace of a matrix is equal to the sum of its eigenvalues, the remaining eigenvalue is equal to the trace of the matrix, so that $\tilde{\lambda}_+=\sum_i \gamma_i$.

Transforming from $\tilde{\lambda}$ back to $\lambda$, we finally find $M-1$ identical eigenvalues $\lambda_- = -\Dc \qq^2$, and one eigenvalue $\lambda_+ = -\Dc \qq^2 - \sum_i \gamma_i$. The latter eigenvalue can become positive, indicating an instability, whenever $\sum_i \gamma_i < 0$ for wave numbers satisfying $\qq^2 < - \Dc^{-1} \sum_i \gamma_i$. The eigenvector corresponding to this unstable eigenvalue takes the form $(u_1,u_2,...,u_M) = (1,\gamma_2/\gamma_1,...,\gamma_M/\gamma_1)u_1$. When rewritten in the original variables, these expressions become Eqs.~1 and 2 in the main text.

\section{Brownian dynamics simulations}

The simulations are performed following the procedure described in Refs.~\citenum{soto14} and \citenum{soto15}, except that we perform them in 3D rather than in 2D.  As we did in the derivation of the continuum theory, we assume that the diffusion of the messenger chemical is much faster than the motion of the active particles. The concentration of chemical around an isolated particle $n$ is therefore given by the steady-state diffusion equation $D \nabla^2 c = 0$, with boundary condition $-D \partial_r c |_{r=R} = \alpha_n /(4 \pi R^2)$ at the surface of the particle expressing the production or consumption of chemical by the particle (taken to be spherical with radius $R$), with activity $\alpha_n$. This boundary-value problem has the solution $c(r)=c_0+\alpha_n/(4\pi Dr)$ where $c_0$ is the chemical concentration at infinity, and implies that each particle carries a `cloud' of chemical with it. In the presence of particle $n$, a second particle $m$ will to lowest order experience a chemotactic velocity $\VV = - \mu_m \nabla c = \frac{\alpha_n \mu_m}{4 \pi D} \frac{\rr_{nm}}{|\rr_{nm}|^3}$, where $\mu_m$ is the mobility of particle $m$, and $\rr_{nm} = \rr_m - \rr_n$ is the relative distance between the two particles. Using a far-field approximation so that the chemical fields induced by each particle can be superimposed on each other (or, equivalently, the velocities induced by each particle added up), the equation of motion for particle $m$ then takes the form 
\begin{equation}
\frac{\mathrm{d} \rr_m}{\mathrm{d} t} = \sum_{n \neq m}  \frac{\alpha_n \mu_m}{4 \pi D}  \frac{\rr_{nm}}{|\rr_{nm}|^3} + \boldsymbol{\xi}_m (t).
\label{eom}
\end{equation}
Here, the subindexes $m,n$ run from 1 to $N_\mathrm{tot}$ where $N_\mathrm{tot}$ is the total number of particles in the system ($N_\mathrm{tot} = N_1 + ... + N_M$ where $M$ is the number of species and $N_i$ the number of particles of species $i$). Eq.~\ref{eom} represents the overdamped Brownian dynamics of particle $m$, with the first term representing the deterministic velocity induced by the presence of all other particles $n\neq m$, and $\boldsymbol{\xi}$ a random velocity representing white noise of intensity $2\Dc$ leading to diffusion of the particle with coefficient $\Dc$.

At each time step, the equations (\ref{eom}) are integrated using a forward Euler scheme. At this stage, in order to account for hard core repulsion between particles, pairs of overlapping particles are reflected  off each other until there are no remaining overlaps, using the ``ellastic collision method'' that has been shown to reproduce (in the limit of small time steps) the correct hard core dynamics \cite{stra99}. We simulate a 3D box with periodic boundary conditions, and interactions are treated using the minimal image convention.

The particle diameter $\sigma=2R$ is used as the basic length scale, and a basic velocity scale can be defined as $V_0 = \alpha_0 \mu_0 /(4 \pi D \sigma^2)$, where $\alpha_0$ and $\mu_0$ are characteristic values of the activity and mobility. Dimensionless activities and mobilities are then defined as $\tilde{\alpha}_i \equiv \alpha_i/\alpha_0$ and $\tilde{\mu}_i \equiv \mu_i/\mu_0$, dimensionless time as $\tau \equiv t V_0 / \sigma$, and the dimensionless noise intensity (equivalent to temperature) as $\tDc \equiv \Dc / V_0 \sigma$. In all simulations performed we use a time step $\delta \tau = 0.001$ and noise intensity $\tDc = 0.01$. Simulations are either run for a total of 1000 particles in a box of size $L=48 \sigma$, or for 4000 particles in a box of size $L=76 \sigma$, leading in both cases to a volume fraction $\phi \approx 0.005$, and initialized with random particle positions. Unless otherwise noted, $\alpha_0$ and $\mu_0$ are chosen so that $|\tilde{\alpha}_1| = |\tilde{\mu}_1| = 1$. The movies are produced at 24 frames/s with a frame taken every 100 time-steps, for a total of 2400 time-steps/s, or an equivalence of 2.4 units of dimensionless time to each second of Movie.

In all simulations reported in the main text, we do not use the Ewald summation method because we are interested in simulating only the finite number of particles in the box, rather than an infinite number of copies of it. Periodic boundary conditions are used instead of a closed box only as a matter of convenience, in order to avoid collisions of fast-moving active molecules or clusters against the walls. Nevertheless, for completeness, we have also performed simulations of binary mixtures using Ewald summation, in which case each particle interacts with an infinite number of copies of itself and all other particles. For these simulations, we used direct Ewald summation \cite{touk03} with minimal image convention, Ewald parameter $\alpha=4/L$, and reciprocal space cutoff $m_\mathrm{max}=3$. We have not observed any qualitative difference in these simulations when compared to those without Ewald summation: in all cases, particles were observed to remain homogeneous, aggregate into static or self-propelled clusters, or separate, under the same conditions as in the simulations without Ewald summation. We also performed simulations with Ewald summation starting from initial conditions given by the final (steady-state) configurations of prior simulations without Ewald summation, and verified that these steady-states (aggregated, separated, self-propelled, etc.) remain stable in the presence of Ewald summation. All in all, the only observable difference we encountered in the presence of Ewald summation was in the separated dense-dilute states formed by mixtures of producer particles, as in Fig. 1(c,center) of the Main Text. For these states, we observed that the ``wall'' of self-repelling particles that forms around the cluster of self-attractive particles (which is a consequence of the periodic boundary conditions) becomes more diffuse in the presence of Ewald summation, but a depletion region still exists around the cluster, see Fig.~S5.

We also note that our equation of motion, Eq.~\ref{eom}, is based only on the far-field contribution of the chemical interactions, and therefore neglects both near-field contributions, as well as hydrodynamic interactions between particles. By a systematic expansion in powers of the inverse separation distance $1/r$, these two corrections can be shown to be of order $1/r^5$ compared to the far-field which is of order $1/r^2$ \cite{varm18,rall19}. Our far-field approximation is therefore fully warranted when particles are far away from each other compared to the particle radius, i.e.~when $r \gg \sigma$. The latter is always true at the onset of the instability and we thus expect our predictions and results for the instability (in particular Eqs.~1 and 2 and Figure 2 of the main text) to be unaffected by near-field and hydrodynamic corrections. As the particles get closer, the near-field and hydrodynamic corrections will become more important and will presumably affect the dynamics of aggregation and growth of the clusters, as well as their internal dynamics. The overall appearance of the clusters, however, is purely determined by the sign of the interactions between particles and we thus expect it to remain similar. Moreover, the neutrality condition in Eq.~3 is determined by the intuitive notion of no net consumption or production of chemical, and is therefore likely to hold even when higher order terms are considered. Near-field effects and hydrodynamic interactions may have a stronger effect on self-propelled states, as the shape-symmetry breaking transition that leads to these states involves rearrangements of the outer layers of the cluster.

\section{Fourier analysis of initial stoichiometry}

In order to compare the stoichiometry at the onset of the instability in simulations against the prediction $\delta \rho_2 = \frac{\mu_2 \rho_{02}}{\mu_1 \rho_{01}} \delta \rho_1$ given by (2) in the Main Text, we perform a Fourier analysis of the simulations. At each timestep in a given simulation, we calculate the coarse-grained concentration fields $\delta \rho_i (\rr)$ for each species, by discretising the simulation box into $4^3 = 64$ bins. This 3-dimensional concentration field is then Fourier-transformed using the Matlab routine ``fftn'' for N-dimensional discrete Fourier transforms, thus obtaining the transformed fields $\delta \rho_i (\qq)$. We then find the optimal ratio $\eta_\mathrm{opt}$ that minimizes the difference $\delta \rho_2 (\qq) - \eta \delta \rho_1 (\qq)$ for each timestep in the least squares sense, obtaining in this way a value for $\eta_\mathrm{opt}$ and a goodness-of-fit indicator $\Gamma$ (which corresponds to the coefficient of determination commonly denoted as $R^2$, but we avoid the latter notation to prevent confusion with the particle radius and to highlight that this coefficient is not necessarily positive) as a function of time. A typical example for the time evolution of the two is shown in Fig.~S2. We see that initially the value of $\Gamma$ is negative, indicating that the hypothesis of a stoichiometric relation $\delta \rho_2 (\qq) = \eta_\mathrm{opt} \delta \rho_1 (\qq)$ is worse than the null hypothesis (the data is better described simply by its average value), as expected given that the initial conditions are random. After some time, however, we observe that $\Gamma$ suddenly increases, becoming positive and approaching 1. We take this rapid increase of $\Gamma$ to be an indicator of the onset of the instability, which we explicitly define as the time at which $\Gamma$ crosses from negative to positive values ($\Gamma=0$). The corresponding value of $\eta_\mathrm{opt}\equiv \delta \rho_2 / \delta \rho_1 |_\mathrm{opt}$ at this point in time is recorded, and the values obtained in this way for all 44 simulations (Table~S1) are shown in Fig.~2(c) of the Main Text.


\clearpage

\section{Supplemental Figures and Table}

 \begin{figure}[h]
 \includegraphics[width=0.5\linewidth]{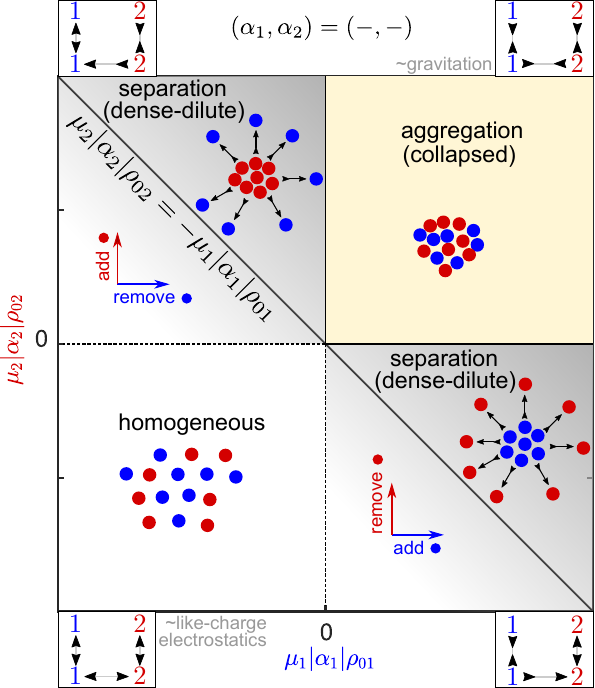}
 \caption{\textbf{Stability diagram for a binary mixture of two consumer species.} The predicted configurations are identical to the case of two producer species discussed in the main text; see Fig.~2(b), except that the sign of the mobilities is switched in this case  $(\mu_1,\mu_2) \to -(\mu_1,\mu_2)$. \label{ext_phasediag}}
 \end{figure}

  \begin{figure}
 \includegraphics[width=0.5\linewidth]{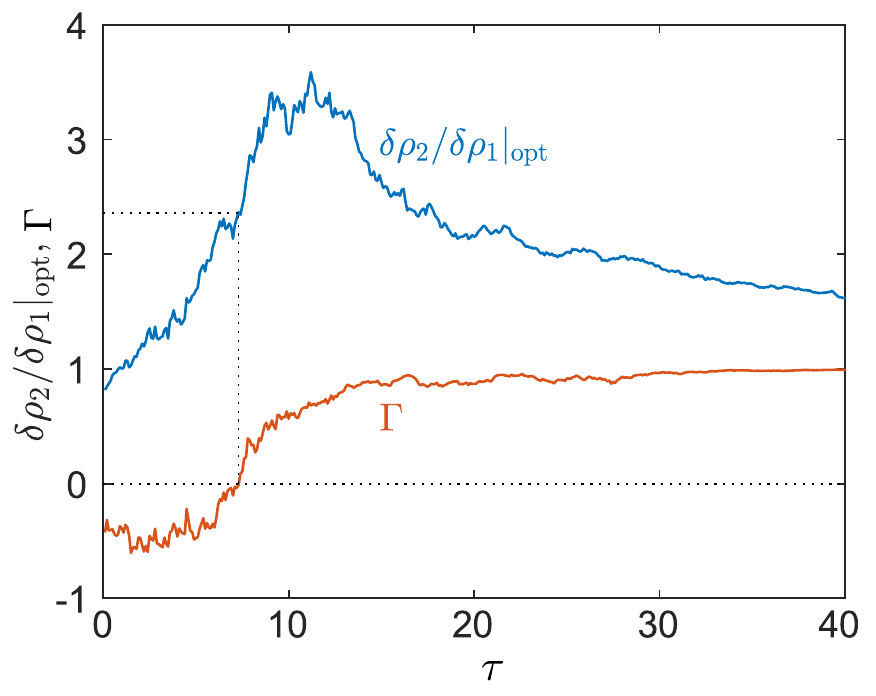}
 \caption{\textbf{Extracting the stoichiometry at the onset of the instability from a given simulation.} We plot the time evolution of the best fit stoichiometry $\eta_\mathrm{opt}\equiv \delta \rho_2 / \delta \rho_1 |_\mathrm{opt}$ and the coefficient of determination $\Gamma$ (goodness-of-fit indicator), obtained from Fourier analysis. The onset of the instability is defined as the moment when $\Gamma$ crosses from negative to positive values ($\Gamma=0$), which is the moment at which the hypothesis of a stoichiometric relation between the fields $\delta \rho_1$ and $\delta \rho_2$ becomes a better fit than the null hypothesis. Each data point in Fig.~2(c) is obtained as a result of this procedure, see also Table~S1. In this case $N_1=N_2=2000$, $\tilde{\alpha}_1 = 1$, $\tilde{\alpha}_2 = -1$, $\tilde{\mu}_1=1$, $\tilde{\mu}_2=2$. \label{ext_fourier}}
  \end{figure}

   \begin{figure}
 \includegraphics[width=1\linewidth]{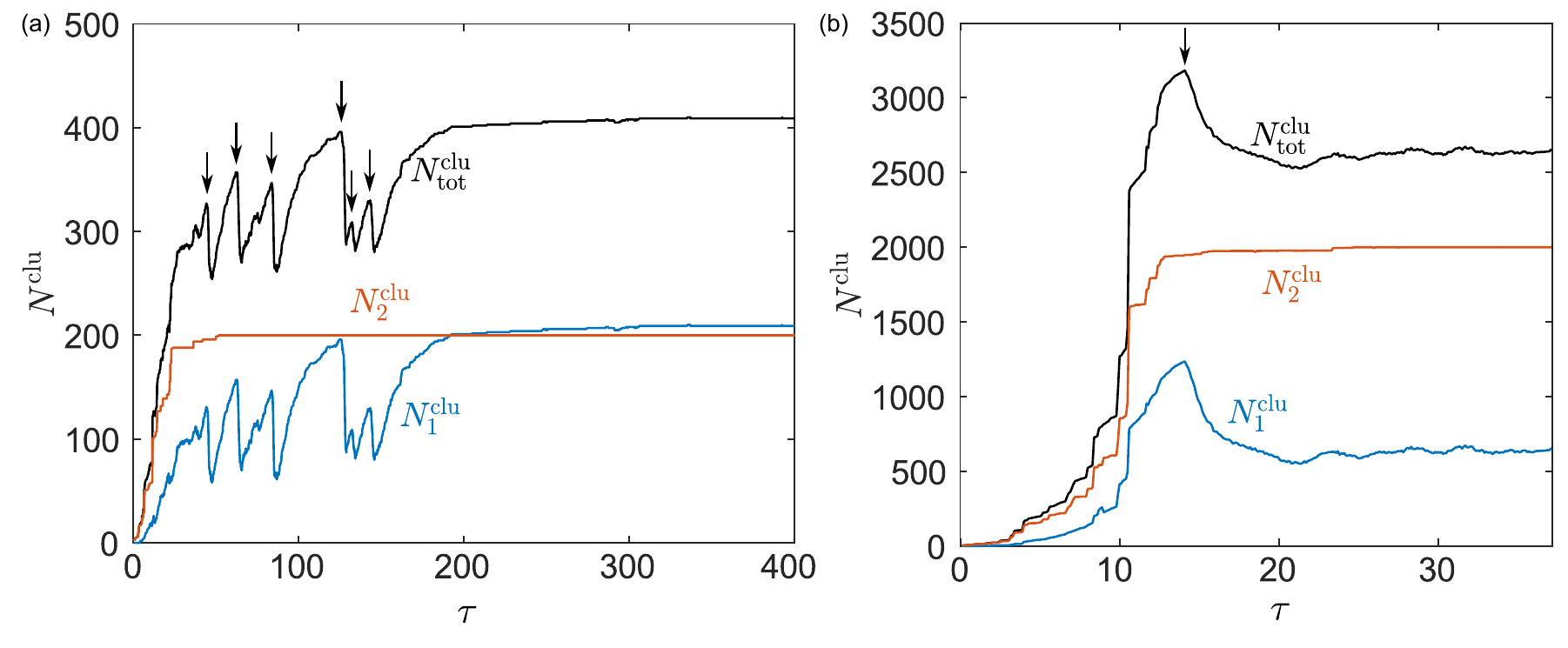}
 \caption{\textbf{Shape instability concomitant with particle ``shedding'' leading to self-propelled macroclusters.} (a) For a small cluster ($N_1=800$, $N_2=200$, $\tilde{\alpha}_1 = 1$, $\tilde{\alpha}_2 = -1$, $\tilde{\mu}_1=1$, $\tilde{\mu}_2=8$), corresponding to movie 8, we see several quick transitions to a transient self-propelled state, which are accompanied by loss of particles of the self-repelling species (here, species 1) marked by arrows. In each case, the lost particles are subsequently recovered, ultimately leading to a stable, static, symmetric ``neutral'' cluster. (b) For a a larger cluster ($N_1=N_2=2000$, $\tilde{\alpha}_1 = 1$, $\tilde{\alpha}_2 = -2$, $\tilde{\mu}_1=1$, $\tilde{\mu}_2=3$), corresponding to movie 9, the lost particles are not recovered and the cluster remains stably in an asymmetric self-propelled state. \label{ext_propelled}}
  \end{figure}
 
  \begin{figure}
 \includegraphics[width=0.5\linewidth]{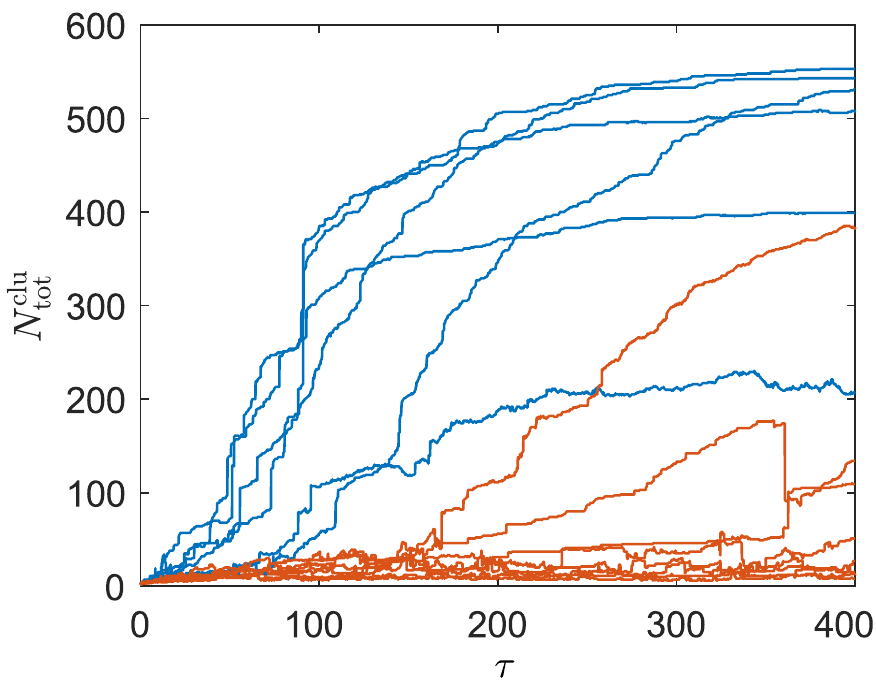}
 \caption{\textbf{Phase separation in highly polydisperse mixtures.} Time evolution of the size of the largest cluster for 14 randomly-generated mixtures of 20 different species each ($N_1=...=N_{20}=50$), with activities and mobilities randomly chosen in the intervals $-2 \leq \tilde{\alpha}_i,\tilde{\mu}_i \leq +2$ for each species. Six of the mixtures (blue) are predicted to be unstable towards phase separation by the instability criterion (1). The eight remaining mixtures (red) are predicted to have a linearly stable homogeneous state, although some of them do appear to phase separate after some time, presumably through a fluctuation-controlled nucleation-and-growth mechanism. \label{ext_polygrowth}}
  \end{figure}
  
    \begin{figure}
 \includegraphics[width=0.75\linewidth]{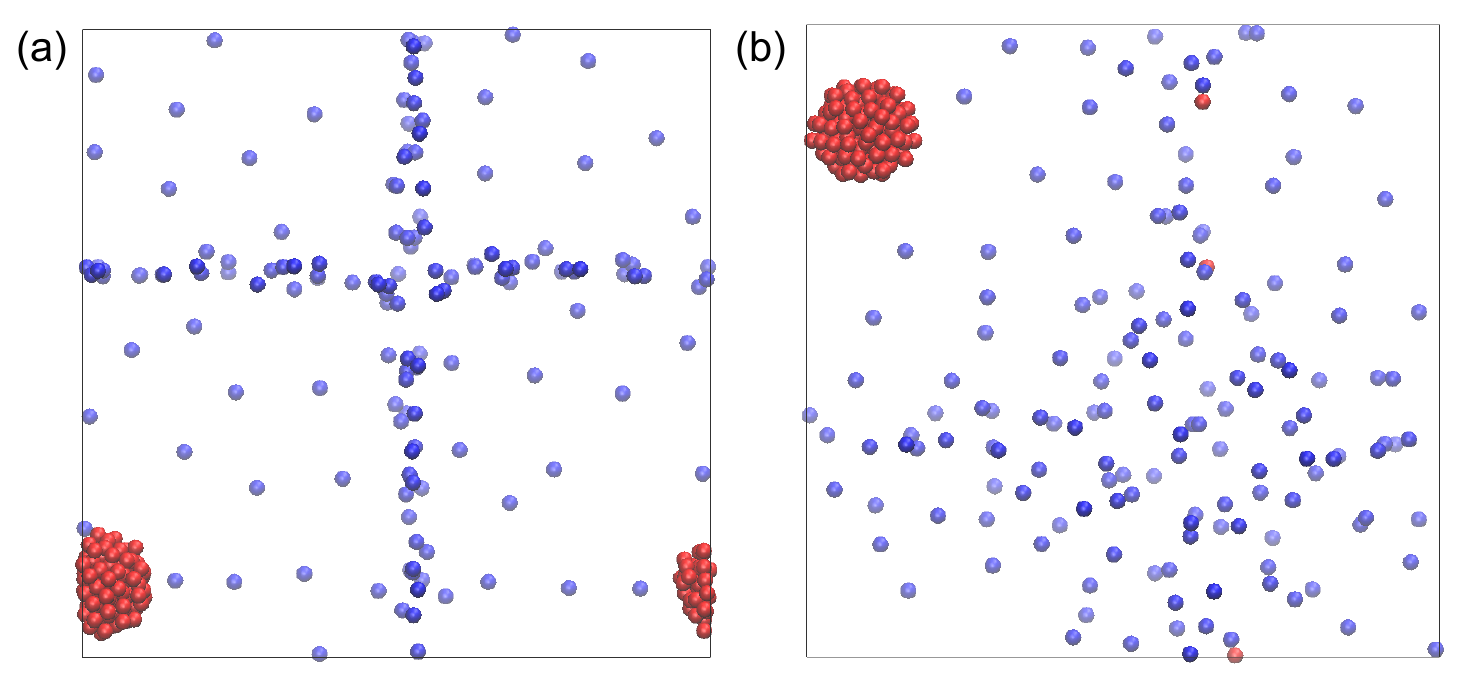}
 \caption{\textbf{Changes on the separated dense-dilute configuration due to long-ranged interactions in a truly periodic lattice.} Steady-state in the absence (a) and presence (b) of Ewald summation, which accounts for the interactions of each particle with all the infinite copies of every other particle and itself in the periodic lattice. The ``walls'' formed by the dilute self-repelling particles (blue) around the dense cluster of self-attracting particles (red) become more diffuse, but there is still an appreciable depletion zone around the cluster. The parameters used are $N_1 = N_2 = 150$, $\tilde{\alpha}_1=\tilde{\alpha}_2=1$, $\tilde{\mu}_1=1$, $\tilde{\mu}_2=-2$, and box size $L = 38 \sigma$ which gives a volume fraction $\phi \approx 0.005$ equivalent to the rest of simulations in this work. The time step was increased to $\delta \tau = 0.01$ to accelerate the simulations. \label{ewald}}
  \end{figure}
 
 \begin{table}
 \caption{Parameters used in the 44 simulations corresponding to Fig.~2(c). \label{ext_stoitable}}
 \begin{tabular}{|c|c|c|c|c|c|c|c|}
 \hline
$\tilde{\alpha}_1$ & $\tilde{\alpha}_2$ & $\tilde{\mu}_1$ & $\tilde{\mu}_2$ & $N_1$ & $N_2$ & $\frac{\mu_2 N_2}{\mu_1 N_1}$ & $\delta \rho_2/\delta \rho_1$ \\
\hline \hline
1 & -1 & 1 & 2 & 2000 & 2000 & 2 & 2.36  \\
1 & -1 & 1 & 3 & 2000 & 2000 & 3 & 4.02 \\
1 & -1 & 1 & 4 & 2000 & 2000 & 4 & 3.85 \\
1 & -2 & 1 & 2 & 2000 & 2000 & 2 & 1.88 \\
1 & -2 & 1 & 3 & 2000 & 2000 & 3 & 3.62 \\
1 & -2 & 1 & 4 & 2000 & 2000 & 4 & 3.83 \\
1 & -1 & 1 & 2 & 1333 & 2667 & 4 & 4.05 \\
1 & -1 & 1 & 3 & 1333 & 2667 & 6 & 7.12 \\
1 & -1 & 1 & 4 & 1333 & 2667 & 8 & 8.45 \\
1 & -1 & 1 & 8 & 800 & 200 & 2 & 2.43 \\
1 & -1 & 1 & 12 & 800 & 200 & 3 & 3.26 \\
1 & -2 & 1 & 4 & 800 & 200 & 1 & 1.15 \\
1 & -2 & 1 & 8 & 800 & 200 & 2 & 1.84 \\
1 & -3 & 1 & 3 & 800 & 200 & 0.75 & 0.78 \\
1 & -3 & 1 & 5 & 800 & 200 & 1.25 & 1.01 \\
1 & -1 & 1 & 2 & 500 & 500 & 2 & 2.17 \\
1 & -1 & 1 & 3 & 500 & 500 & 3 & 3.91 \\
1 & -1/2 & 1 & 4 & 500 & 500 & 4 & 5.43 \\
1 & -1/2 & 1 & 6 & 500 & 500 & 6 & 7.91 \\
1 & -1/3 & 1 & 6 & 500 & 500 & 6 & 6.19 \\
1 & -1 & 1 & 2 & 500 & 500 & 2 & 2.83 \\
1 & -1 & 1 & 4 & 500 & 500 & 4 & 4.67 \\
1 & -1 & 1 & 6 & 500 & 500 & 6 & 6.19 \\
1 & -1 & 1 & 8 & 500 & 500 & 8 & 8.05 \\
1 & -1 & 1/8 & 1 & 500 & 500 & 8 & 8.65 \\
1 & -1 & 1/4 & 1 & 500 & 500 & 4 & 4.42 \\
1/2 & -1 & 1/8 & 1 & 500 & 500 & 8 & 7.38 \\
1/3 & -1 & 1/2 & 1 & 500 & 500 & 3 & 2.18 \\
1 & -1 & 1/8 & 1 & 800 & 200 & 2 & 1.71 \\
1 & -1 & 1/12 & 1 & 800 & 200 & 3 & 3.85 \\
1/2 & -1 & 1/4 & 1 & 800 & 200 & 1 & 1.37 \\
1/2 & -1 & 1/8 & 1 & 800 & 200 & 2 & 1.99 \\
1/3 & -1 & 1/2 & 1 & 800 & 200 & 0.75 & 0.8 \\
1/3 & -1 & 1/4 & 1 & 800 & 200 & 1.25 & 1.48 \\
1 & -1/3 & 1 & 6 & 500 & 500 & 6 & 5.78 \\
1 & -1/3 & 1 & 6 & 500 & 500 & 6 & 5.44 \\
1 & -1/3 & 1 & 6 & 500 & 500 & 6 & 6.94 \\
1 & -1 & 1 & 3 & 500 & 500 & 3 & 3.68 \\
1 & -1 & 1 & 2 & 500 & 500 & 2 & 2.59 \\
1 & -3 & 1 & 1 & 500 & 500 & 1 & 1.08 \\
1 & -1 & 1 & 1 & 400 & 600 & 1.5 & 2.42 \\
1 & -3 & 1 & 4 & 800 & 200 & 1 & 1.23 \\
1 & -1/3 & 1 & 4 & 400 & 600 & 6 & 6.27 \\
1 & -1 & 1 & 1.25 & 500 & 500 & 1.25 & 1.79 \\
\hline
 \end{tabular}
 \end{table}

  \clearpage
 
 \section{Description of the movies}
 
 \begin{itemize}
 
\item \textbf{Movie 1:} Mixture of producer and consumer species that remains homogeneous with formation of small molecules (in particular, self-propelling dimers); see Figs.~1(a), left, and 2(a). Parameters used $N_1=N_2=500$, $\tilde{\alpha}_1 = 1$, $\tilde{\alpha}_2 = -1$, $\tilde{\mu}_1=1$, $\tilde{\mu}_2=1/2$.
  
\item \textbf{Movie 2:} Mixture of producer and consumer species showing aggregation into a neutral static cluster that coexists with a dilute phase, see Figs.~1(a), centre, and 2(a). Parameters used $N_1=N_2=500$, $\tilde{\alpha}_1 = 1$, $\tilde{\alpha}_2 = -1/2$, $\tilde{\mu}_1=1$, $\tilde{\mu}_2=6$.
  
\item \textbf{Movie 3:} Mixture of producer and consumer species showing separation into two collapsed clusters, see Figs.~1(a),right and 2(a). Parameters used $N_1=N_2=2000$, $\tilde{\alpha}_1 = 1$, $\tilde{\alpha}_2 = -1$, $\tilde{\mu}_1=-1$, $\tilde{\mu}_2=2$.
  
\item \textbf{Movie 4:} Mixture of two producer species that remains homogeneous; see Figs.~1(c), left, and 2(b). Parameters used $N_1=N_2=500$, $\tilde{\alpha}_1 = 1$, $\tilde{\alpha}_2 = 1$, $\tilde{\mu}_1=1$, $\tilde{\mu}_2=1/2$.
  
\item \textbf{Movie 5:} Mixture of two producer species that shows aggregation into a single collapsed cluster; see Figs.~1(c), right, and 2(b). Parameters used $N_1=N_2=2000$, $\tilde{\alpha}_1 = 1$, $\tilde{\alpha}_2 = 1$, $\tilde{\mu}_1=-1$, $\tilde{\mu}_2=-2$.
  
\item \textbf{Movie 6:} Mixture of two producer species that shows separation into a dense phase and a dilute
phase that are pushed away from each other; see Figs.~1(c), centre, and 2(b). Parameters used $N_1=N_2=500$, $\tilde{\alpha}_1 = 1$, $\tilde{\alpha}_2 = 1$, $\tilde{\mu}_1=1$, $\tilde{\mu}_2=-2$.
  
\item \textbf{Movie 7:} Mixture of two producer species for which we predict a linearly stable homogeneous state (see Fig.~2(b)), but still shows slow separation between dense and dilute phases, presumably formed by fluctuation-controlled nucleation-and-growth. Parameters used $N_1=N_2=500$, $\tilde{\alpha}_1 = 1$, $\tilde{\alpha}_2 = 1$, $\tilde{\mu}_1=1$, $\tilde{\mu}_2=-1/2$.
  
\item \textbf{Movie 8:} Mixture of producer and consumer species showing several quick transitions between a static cluster and a self-propelled cluster state, involving the quick shedding and slow recovery of the self-repelling (blue) particles, see Fig.~S3(a). Finally, the static state remains stable. Parameters used $N_1=800$, $N_2=200$, $\tilde{\alpha}_1 = 1$, $\tilde{\alpha}_2 = -1$, $\tilde{\mu}_1=1$, $\tilde{\mu}_2=8$.
  
\item \textbf{Movie 9:} Mixture of producer and consumer species showing a transition into a stable self-propelled cluster state, involving the quick shedding the self-repelling (blue) particles; see Fig.~1(b) and Fig.~S3(b). Parameters used $N_1=N_2=2000$, $\tilde{\alpha}_1 = 1$, $\tilde{\alpha}_2 = -2$, $\tilde{\mu}_1=1$, $\tilde{\mu}_2=3$.
  
\item \textbf{Movie 10:} Phase separation induced by a small amount of an active ``doping agent''. Adding a small amount of a tird species makes an otherwise homogeneous mixture (cf.~movie 1) unstable towards phase separation, see also Fig.~3. Parameters used $N_1=N_2=500$, $N_3=50$, $\tilde{\alpha}_1 = 1$, $\tilde{\alpha}_2 = -1$, $\tilde{\alpha}_3 = -5$, $\tilde{\mu}_1=1$, $\tilde{\mu}_2=1/2$, $\tilde{\mu}_3=2$.
  
\item \textbf{Movie 11:} Highly polydisperse mixture (20 species, with 50 particles of each species for a total of 1000 particles) with randomly generated activities and mobilities, not satisfying the instability criterion (1), that remains homogeneous.
  
\item \textbf{Movie 12:} Highly polydisperse mixture (20 species, with 50 particles of each species for a total of 1000 particles) with randomly generated activities and mobilities, satisfying the instability criterion (1), that shows phase separation into a cluster and a dilute phase. The dilute phase shows formation of small active molecules (e.g. dimers that align with and follow the cluster).
 
 \end{itemize}